# Expected RIP: Conditioning of The Modulated Wideband Converter

Moshe Mishali and Yonina C. Eldar
Department of Electrical Engineering
Technion — Israel Institute of Technology
Haifa, Israel 32000
Email: moshiko@tx.technion.ac.il, yonina@ee.technion.ac.il

*Abstract*—The sensing matrix of a compressive system impacts the stability of the associated sparse recovery problem. In this paper, we study the sensing matrix of the modulated wideband converter, a recently proposed system for sub-Nyquist sampling of analog sparse signals. Attempting to quantify the conditioning of the converter sensing matrix with existing approaches leads to unreasonable rate requirements, due to the relatively small size of this matrix. We propose a new conditioning criterion, named the expected restricted isometry property, and derive theoretical guarantees for the converter to satisfy this property. We then show that applying these conditions to popular binary sequences, such as maximal codes or Gold codes, leads to practical rate requirements.

*Index Terms*—Compressed sensing, expected restricted isometry property, modulated wideband converter, multiband sampling.

## I. INTRODUCTION

Signal dimensions in today's applications are growing faster than technology capabilities. The Nyquist rate of analog wideband signals, for example, already exceeds the conversion rate of existing devices. The modulated wideband converter (MWC) is a recent sub-Nyquist sampling system which exploits frequency sparsity to reduce the conversion rate [1]. Figure 1 depicts a block diagram of the converter, which is further described in Section II. The key idea underlying the MWC is that if the signal is periodically-modulated prior to sampling, then the sampling rate can be substantially reduced with respect to the Nyquist rate. The MWC consists of simple mixers and lowpass filters which are easy to implement. Reconstruction of the analog input from the MWC samples is nonlinear when the frequency support of the signal is unknown; a concrete recovery algorithm is detailed in [1].

In this paper, we study the conditioning of the MWC sampling operator, or equivalently the ability to recover the input in a stable manner. Mathematically, the main nonlinear step boils down to solving for the sparsest solution of a linear underdetermined linear system – a well-studied subject within the compressed sensing (CS) literature. The stability of sparse recovery is dictated by the conditioning of the sensing matrix, which in the MWC configuration depends on the parameters of the system, such as the number of channels and the choice of mixing waveforms. In Section III, we apply known CS results in order to quantify the required number of sampling channels that ensure stability. Unfortunately, these results lead to an unreasonable system size. The relatively small sensing matrix of the MWC is responsible for this behavior, since theoretical constants that are often ignored in large-scale problems have noneligible contribution otherwise.

Going from theory to practice, in Section IV, we aim at practical conditioning guarantees for the MWC, namely for a small number of channels, as the empirical evidence in [1] demonstrates. To achieve this goal, we first introduce a new stability criterion, termed the

This work was supported in part by the Israel Science Foundation under Grant no. 1081/07 and by the European Commission in the framework of the FP7 Network of Excellence in Wireless COMmunications NEWCOM++ (contract no. 216715).

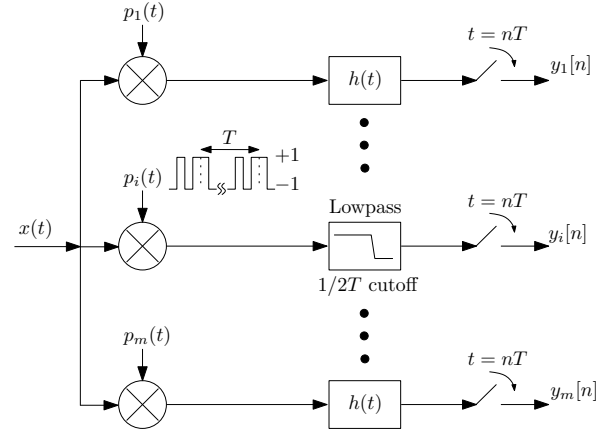

**Fig. 1:** The modulated wideband converter consists of $m$ parallel channels, which mix the input by periodic waveforms. The mixed signal is then lowpass filtered and sampled at a low rate.

expected restricted isometry property (ExRIP), which quantifies the stability of a given (deterministic) sensing operator when applied to random sparse signals. The ExRIP extends on two related definitions: the RIP [2] which assumes no randomness and in general cannot be computed in polynomial-time, and the statistical-RIP [3]. The latter uses a partial random model in which the sensing matrix is deterministic while the nonzero locations are random, and it is limited to matrices with stringent structure, which the MWC, for example, does not posses. The ExRIP relaxes the requirements on the sensing operator by considering a fully random signal model. Our main contribution is in proving that the MWC has the ExRIP when wisely selecting periodic mixing waveforms. Specifically, we show that popular binary sequences, such as maximal codes or Gold codes [4], are adequate candidates that yield reasonable requirements on the system size.

## II. THE MODULATED WIDEBAND CONVERTER

In this section, we begin by describing the sensing mechanism of the MWC. We then formulate our recovery problem and discuss the role of conditioning.

### A. Sensing

The MWC consists of an analog front-end with $m$ channels. In the $i$th channel, the input signal $x(t)$ is multiplied by a periodic waveform $p_i(t)$, lowpass filtered, and then sampled at rate $1/T$. In this paper, we study a simplified version of the converter, as depicted in Fig. 1, in which the sampling interval $T$ equals the period of the waveforms $p_i(t)$. In addition, $p_i(t)$ are chosen as sign alternation waveforms, such that each period $T$ consists of $M$ intervals of length $T/M$ each, and $p_i(t) = \pm 1$ on each such interval. This basic configuration is sufficient for studying the fundamental theoretical trade-off between rate and stability; other configurations with practical advantages are detailed in [1].

The MWC was studied in [1] mainly for multiband analog signals. The support of a multiband signal $x(t)$ resides within $N$ frequency intervals, or bands, such that the width of each band does not exceed $B$ Hz. The band positions are arbitrary and in particular unknown in advance. For example, in communication $N$ represents the number of concurrent transmissions and $B$ is specified by the transmission techniques in use. We note that sub-Nyquist sampling is one of the appealing properties of the MWC, though it can also be used for conventional Nyquist sampling with the proper number of channels.

The MWC sensing relies on the following key observation. The mixing operation scrambles the spectrum of $x(t)$ such that the baseband frequencies that reside below the filter cutoff $1/2T$, contain a mixture of the spectral contents from the entire Nyquist range. The periodicity of each waveform $p_i(t)$ ensures that the mixture has a specific nature – aliases at $1/T$ frequency spacing. Whilst aliasing is often considered as an undesired effect, here it is deliberately utilized to shift various frequency regions to baseband, simultaneously. In the basic configuration, we choose the rate $1/T = B$ and the length $M$ of the sign patterns $p_i(t)$ is set to the compression ratio, namely the integer that is closest to the quotient of the Nyquist rate by $1/T$.

### B. Reconstruction

The recovery of $x(t)$ from the digital sequences $y_1[n], \ldots, y_m[n]$ consists of two steps which both exploit the sparse nature of the multiband spectrum. First, the spectral support is determined, and then the signal is recovered from the samples by a closed-form expression. The support recovery involves a series of digital computations, which are grouped together under the Continuous-to-Finite (CTF) block [1], [5]. As the name hints, the CTF allows to treat the resulting continuous recovery problem efficiently, by inferring the support from a small-size finite program. In the noiseless scenario, once the support is found by the CTF block, the input signal $x(t)$ is perfectly recovered. When noise is present, it may impact both the digital support recovery and the actual continuous reconstruction. We refer the reader to [1], [5] for a detailed discussion on the recovery process.

One of the elements of the CTF, which is our main focus here, is solving an underdetermined linear system for the sparsest solution matrix; also known as multiple measurement vectors (MMV) problem in the CS literature. The CTF block generates the MMV system

$$\mathbf{V} = \mathbf{A}\mathbf{U}, \quad (1)$$

where $\mathbf{V}$ is an $m \times r$ matrix that is computed from the given sequences $y_1[n], \ldots, y_m[n]$, with $r > 0$ some positive integer. The goal is to find an $M \times r$ matrix $\mathbf{U}$ with as few nonzero rows as possible. The nonzero rows in the sparsest $\mathbf{U}$ indicate the unknown support of $x(t)$ [1]. The matrix $\mathbf{A}$ represents the sensing operator

$$\mathbf{A} = \mathbf{SFD}, \quad (2)$$

where $\mathbf{S}$ is an $m \times M$ matrix, whose $i$th row contains the sign pattern of the $i$th waveform $p_i(t)$. In the basic configuration $\mathbf{F}$ is the $M$-square DFT matrix (up to a column permutation). The matrix $\mathbf{D}$ is diagonal and accounts for the decay of the Fourier transform of $p_i(t)$ at high frequencies. The decay has to be compensated by analog means but this subject is beyond the current scope. For our purposes $\mathbf{D}$ can be ignored, since the nonzero location set $\text{supp}(\mathbf{u}) = \text{supp}(\mathbf{D}\mathbf{u})$ for any vector $\mathbf{u}$. Therefore, from this point on we focus on the sensing part $\mathbf{SF}$ of the matrix $\mathbf{A}$.

A large body of CS literature studies sparse recovery problems, such as (2), with either $r \geq 1$. It is well-known that finding the sparsest $\mathbf{U}$ is NP-hard in general. Fortunately, there are many sub-optimal polynomial-time algorithms that yield the exact solution under different conditions on the sensing matrix. Typical recovery conditions of CS algorithms require the number $m$ of rows to be proportional to the cardinality $K$ of the nonzeros support. A logarithmic dependency on the number $M$ of columns is also necessary for stable recovery [6]. In our setting, these conditions translate to a requirement on the number of parallel channels in the MWC. Obviously, we would like to have theoretical recovery guarantees with a reasonable number of channels, since those are implemented in hardware.

The simulations in [1] show 97% accurate support recovery rate on extensive sets of band locations. In the simulations, 3 concurrent transmissions, each of width $B = 50$ MHz, were generated with additive noise, where a Nyquist rate of 10 GHz defined the wideband input range. The MWC system used $m = 40$ channels with $M = 195$ length sign-patterns and rate $1/T \approx B$, which implies 80% rate saving with respect to the Nyquist rate. In this setting, there are 12 nonzero rows at most in $\mathbf{U}$ but due to the conjugate symmetry of the Fourier transform, it amounts to sparse recovery with $K = 6$ nonzeros. These numbers will serve us as a gold standard. As mentioned earlier, in practice the number of channels $m$ can be substantially reduced when using other configurations of the MWC.

In the next section, we study existing conditions from the CS literature and show that they require a prohibitively large number of channels. Then, in Section IV we show how a wise selection of the sign patterns in $p_i(t)$ leads to conditioning guarantees with a small number of channels.

### III. APPLYING KNOWN RECOVERY GUARANTEES

In what follows, $\mathbf{\Phi}$ denotes an arbitrary matrix of size $m \times M$ with $m < M$, and $\mathbf{u}$ is an unknown $K$-sparse vector, with no more than $K$ nonzeros. The goal is to ensure that the recovery of $\mathbf{u}$ from the underdetermined measurement $\mathbf{v} = \mathbf{\Phi}\mathbf{u}$ is well-conditioned. We study the scenario (1) with $r = 1$ but comment that for $r > 1$, which is the typical MMV dimensions that the CTF generates, slightly better results can be obtained. In addition, we relate all conditions to the convex recovery method

$$\min_{\mathbf{u}} \|\mathbf{u}\|_1 \text{ s.t. } \|\mathbf{v} - \mathbf{A}\mathbf{u}\|_2^2 \leq \epsilon. \quad (3)$$

Program (3) is known as basis pursuit (BP) for $\epsilon = 0$. BP denoising refers to the case $\epsilon > 0$. The quadratic constraint is sometimes regularized and merged into the objective.

### A. Coherence-based Guarantees

The coherence of a matrix $\mathbf{\Phi}$ is defined as

$$\mu = \max_{i \neq j} \frac{|\langle \mathbf{\Phi}_i, \mathbf{\Phi}_j \rangle|}{\|\mathbf{\Phi}_i\| \|\mathbf{\Phi}_j\|}, \quad (4)$$

where the subscript $\mathbf{\Phi}_i$ denotes the $i$th column. The coherence $\mu$ can be efficiently computed for any given matrix. A well known CS result [7, Th. 7] is that if $K \leq 0.5(1 + 1/\mu)$, then any $K$-sparse vector $\mathbf{u}$ is perfectly recovered by BP.

Another result by Tropp [8, App. IV-A] shows that if $K \leq 1/3\mu$, and the measurement $\mathbf{\Phi}\mathbf{u}$ is contaminated by Gaussian white noise with covariance $\sigma^2\mathbf{I}$, then the support of $\mathbf{u}$ can be recovered with high probability using the BP denoising program.

Candès and Plan [9, Th. 1.2] considered the noisy model in which the support $\text{supp}(\mathbf{u})$ is drawn uniformly at random among all possible choices, and the nonzero values have amplitudes $|u_i| > (6 + \sqrt{2})\sigma\sqrt{2\log M}$ and random signs. Under the conditions $\mu < c/\log M$ and $K \leq cM/\|\mathbf{\Phi}\|^2 \log M$, the BP denoising is proved to recover the support with high probability.

In Table I, we numerically evaluate the recovery guarantees based on these bounds. The "best-case" column indicates the smallest dimensions $m, M$ of $\mathbf{S}$, such that the theoretical requirements are satisfied for the given $K$. For example, for $K = 2$ the bound of [7] is satisfied with $\mathbf{S}$ of dimensions $300 \times 2000$ at least. We selected comparable dimensions for $\mathbf{S}$ and slightly adjusted $K$ where required. The best matrix $\mathbf{S}$, in the sense of lowest $\mu(\mathbf{SF})$, was chosen out of 100 realizations with $\pm 1$ entries that were drawn independently with equal probability. We could not verify the exact value of the constant $c$ from the statement or the proof of [9]. For coherence results the probability $p = 1$ in the table means no randomness in both $\mathbf{\Phi}$ and $\mathbf{u}$. The MWC column assumes $M = 195, K = 12$ and the value of $m$ that is required to satisfy the relevant bounds is displayed. As before, the instance $\mathbf{S}$ which gives the lowest value for $m$ is used.

As the table shows, in small-scale problems the coherence-based guarantees require an unreasonable number of channels, leading to a sampling rate that exceeds Nyquist by orders of magnitude. Evidently, the constants can be very significant.

We next examine conditioning guarantees that measure the correlation between large subsets of columns, rather than pairs as in (4). Obviously, these conditions can better predict the conditioning of $\mathbf{\Phi}$ on sparse vectors, but as it turns out they are harder to compute for a given matrix.

*B. RIP Guarantees*

Candès et. al. [2] introduced the restricted isometry property (RIP) as a standard tool for analyzing sensing matrices. A matrix $\mathbf{\Phi}$ is said to have the RIP with isometry constant $\delta_K$, if $0 \leq \delta_K < 1$ is the smallest number such that

$$(1 - \delta_K)\|\mathbf{u}\|^2 \leq \|\mathbf{Au}\|^2 \leq (1 + \delta_K)\|\mathbf{u}\|^2 \quad (5)$$

holds for *all* $K$-sparse vectors $\mathbf{u}$ [2]. The RIP quantifies how well $\mathbf{\Phi}$ preserves the norm of sparse vectors. If $\delta_{2K} < 1$ then every $K$-sparse vector $\mathbf{u}$ is uniquely determined by $\mathbf{\Phi u}$. Furthermore, if $\delta_{2K} < \sqrt{2} - 1$ then BP recovers $\mathbf{u}$ exactly [10]. In the presence of noise, the same condition ensures that BP denoising recovers the signal up to a small bounded error [10].

The main drawback of RIP conditions is that the isometry constant $\delta_{2K}$ cannot be computed in polynomial time for an arbitrary deterministic matrix. The common workaround is to consider random matrix ensembles, e.g. when the entries of $\mathbf{\Phi}$ are drawn from the Gaussian, or the Bernoulli distributions. A relevant result appears in [6] based on [11]:

*Theorem 1:* Let $\mathbf{\Phi}$ be an $m \times M$ matrix generated by drawing entries from an appropriately scaled sub-Gaussian distribution. If

$$m \geq \frac{2}{c\delta_K}\left(\ln(2L) + K \ln\left(\frac{12}{\delta_K}\right) + t\right), \quad (6)$$

where $L = \binom{M}{K}$ and $c$ is a distribution-dependent constant, then, with probability at least $p = 1 - e^{-t}$, $\mathbf{\Phi}$ has the RIP.

In our setting, if $\mathbf{S}$ is random with equali-likely $\pm 1$ entries, and (6) is satisfied, then it has the RIP. The constant $c = 7/18$ in this setting [11]. Since $\mathbf{F}$ is a unitary matrix, the compounded sensing matrix $\mathbf{SF}$ has the same RIP constant as $\mathbf{S}$ [11]. In Table I we calculate the typical setting in which the bound (6) is satisfied for $\delta_{2K} = \sqrt{2} - 1$ and $p = 0.97$. As before, the theoretical requirements seem pessimistic.

Besides the requirement for a large number of sampling channels, there is a delicate logical issue in the above inference. Theorem 1 predicts the RIP property of a random matrix. In practice, the entries of $\mathbf{S}$ are fixed to the specific sign patterns that are realized in the system. Notice also the different meaning of the probability; in Theorem 1 it refers to instances of $\mathbf{S}$, while in empirical simulations the recovery rate refers to signal realizations with fixed $\mathbf{S}$.

A recent approach in CS considers deterministic matrices by switching the role of randomness from the sensing operator to the signal model. This framework conforms with the conventional Bayesian approach in signal processing, and naturally fits simulation methodology. In the next section we quote results of this kind.

*C. Statistical-RIP Guarantees*

Calderbank et. al. [3] proposed the Statistical RIP (StRIP) as an alternative tool for quantifying sensing matrices. A matrix $\mathbf{\Phi}$ has the StRIP($K, \delta_K, p$) if (5) is satisfied with probability at least $p$ for a $K$-sparse vectors $\mathbf{u}$, whose support is uniformly drawn from all possible choices (the nonzero values are arbitrary). Calculating the StRIP for an arbitrary matrix $\mathbf{\Phi}$ is not easier than RIP computations. However, structured matrices can greatly simplify the calculations. In [3] the authors consider matrices, whose columns form a closed-group under element-wise multiplication. In addition, it is assumed that the rows of $\mathbf{\Phi}$ are orthogonal and each sums to zero. Under these hypothesis they prove that the StRIP is satisfied with

$$\frac{K-1}{M-1} < \delta_K < 1 \qquad p \geq 1 - \frac{\frac{2K}{m} + \frac{2K+7}{M-3}}{\left(\delta_K - \frac{K-1}{M-1}\right)^2}. \quad (7)$$

This result is however not useful for the MWC, since the columns of $\mathbf{SF}$ do not form the required group, and this property is essential for the arguments in [3]. Even when ignoring this issue, the required dimensions $m, M$ are high. In the MWC setting $m = 150$ channels give probability $p = 0$ in (7).

More recently, Gan et. al. [12] studied StRIP guarantees for $\mathbf{\Phi}$ with unit-norm columns and zero-sum rows. They showed that the StRIP is satisfied with

$$\frac{1}{M-1} < \delta_K \qquad p \geq 1 - 2\exp\left(-\frac{\left(\delta_K - \frac{1}{M-1}\right)^2}{16\mu^2 K}\right), \quad (8)$$

where as before $\mu$ is the coherence of $\mathbf{\Phi}$. See Table I for evaluation of this result.

Tropp [13, Th. 12] also provides guarantees on the conditioning of a deterministic matrix. We bring this result using StRIP terminology. A matrix $\mathbf{\Phi}$ with unit-norm columns satisfies the StRIP with probability at least $p = 1 - (K/2)^{-t}$, for some $t \geq 1$, if

$$\sqrt{144\mu^2 K t \log(K/2 + 1)} + \frac{2K}{M}\|\mathbf{\Phi}\|^2 \leq e^{-1/4}\delta_K. \quad (9)$$

Additional conditions are used to bound the probability that BP recovers $\mathbf{u}$ exactly [13, Th. 14]. Observe the table for the applicability of this bound. Note that both [12], [13] require $\mathbf{\Phi}$ to have unit-norm columns. Therefore we applied them on $\mathbf{\Phi} = \mathbf{SF}/\sqrt{mM}$ which approximately satisfies this requirement.

The attempt to derive practical recovery guarantees for the MWC, which are based on existing results for general structured $\mathbf{\Phi}$, is perhaps sentenced to fail; It does not take into account the specific structure of the sensing matrix $\mathbf{SF}$. The next section capitalizes on this structure in order to reduce the requirements on the number of sampling channels.

IV. CONDITIONING OF THE MWC

*A. The ExRIP*

We begin with defining a StRIP-like property, which accounts for randomness in the nonzero values.

TABLE I: Recovery guarantees for the MWC

| | Conditioning measure | "Best-case" setting | | | | MWC Setting (see text) | Support | Values | Noise | Remarks/Issues |
|---|---|---|---|---|---|---|---|---|---|---|
| | | $m$ | $M$ | $K$ | $p$ | | | | | |
| Coherence | Donoho-Elad [7] | 300 | 2000 | 2 | 1 | $m \geq 4230$ | D | D | - | |
| | Tropp [8] | 300 | 2000 | 1 | 1 | $m \geq 9540$ | D | D | + | |
| | Candès-Plan [9] | | n/a | | | n/a | R | Random signs | + | unspecified constant |
| RIP | Blumensath+ [6], [11] | 700 | 5000 | 3 | 0.95 | $m \geq 950$ | D | D | - | random $\mathbf{S}$ |
| StRIP | Calderbank+ [3] | 700 | 5000 | 3 | 0.93 | $m = 150, p = 0$ | R | D | - | columns form a group |
| | Gan+ [12] | 500 | 3000 | 1 | 0 | n/a | R | D | - | unit-norm columns |
| | Tropp [13] | 1000 | 2000 | 3 | 0.01 | n/a | R | D | - | unit-norm columns |
| ExRIP | This paper | 80 | 511 | 12 | 0.94 | $m \geq 40$ | R | R | - | $\mathbf{\Phi} = \mathbf{SF}$, theory |
| | Simulations [1] | 40 | 195 | 12 | 0.96 | $m \geq 40$ | R | R | + | $\mathbf{\Phi} = \mathbf{SF}$, simulations |

D=deterministic, R=random, +/-=with/without noise, n/a=not applicable

*Definition 1:* A matrix $\mathbf{\Phi}$ has the expected restricted isometry property (ExRIP), if (5) holds with probability at least $p$ for $K$-sparse random vectors $\mathbf{u}$ whose support is uniformly distributed and whose nonzeros are i.i.d random variables.

The ExRIP involves several constants: the sparsity level $K$, the isometry constant $\delta_K$, the probability $p$, and finally the distribution from which the nonzeros are drawn. Both ExRIP and StRIP assume random support. However, the ExRIP adds another layer of randomness in the nonzero values. On the one hand, the ExRIP is mathematically weaker since worst-case signal values that are unlikely to encounter are averaged out. On the other hand, since the random support assumption anyway excludes worst-case scenarios it makes sense to consider random values as well. Moreover, random signal values conform with the conventional Bayesian framework. Another advantage of the ExRIP is in simplifying complicated expressions that otherwise require $\mathbf{\Phi}$ to have a stringent structure as in [3]. Compare between the StRIP [3] and the ExRIP [14] proofs for instance. Besides these reasons, applying the StRIP results in high system dimensions, whereas as Table I shows the ExRIP is the only measure that leads to a reasonable number of channels.

Our goal is to prove that the sensing matrix $\mathbf{SF}$ has the ExRIP with high probability, for the MWC dimensions. To achieve this goal, we characterize the quality of a given set of sign patterns as follows. The correlation of the rows is captured by

$$\boldsymbol{\alpha}(\mathbf{S}) = \frac{1}{(mM)^2} \sum_{i,k=1}^{m} (\mathbf{S}_i^T \mathbf{S}_k)^2, \quad (10)$$

where the subscripts indicate the relevant rows of $\mathbf{S}$. Note that $\boldsymbol{\alpha}(\mathbf{S})$ resembles the coherence $\mu$ with two distinguishing properties: the coherence is computed over the columns of $\mathbf{SF}$, while $\boldsymbol{\alpha}(\mathbf{S})$ involves the rows of $\mathbf{S}$ only. In addition, $\mu$ involves maximization while $\boldsymbol{\alpha}(\mathbf{S})$ computes sums of squares. Another quality of interest is the total power of all auto- and cross-correlation functions, as measured by

$$\boldsymbol{\beta}(\mathbf{S}) = \frac{1}{m^2 M^3} \sum_{i,k=1}^{m} \|\mathbf{S}_i \odot \mathbf{S}_k\|^2. \quad (11)$$

Here $\mathbf{S}_i \odot \mathbf{S}_k$ stands for cyclic convolution. We will also need

$$\boldsymbol{\gamma}(\mathbf{S}) = \frac{1}{(mM)^2} \sum_{i,k=1}^{m} (\mathbf{S}_i^T \mathbf{S}_k^-)^2, \quad (12)$$

where for any vector $\mathbf{a}^-[n] = \mathbf{a}[-n], n = 0, \ldots, M-1$ under the convention that the indices are module $M$.

The quality measures are bounded below and above by:

$$\frac{1}{m} \leq \boldsymbol{\alpha}(\mathbf{S}) \leq 1 \qquad \frac{1}{M} \approx \frac{2m-1}{2mM-1} \leq \boldsymbol{\beta}(\mathbf{S}), \boldsymbol{\gamma}(\mathbf{S}) \leq 1. \quad (13)$$

Orthogonal rows achieve the lowest $\boldsymbol{\alpha}(\mathbf{S})$ [14]. The lower bound on $\boldsymbol{\beta}(\mathbf{S}), \boldsymbol{\gamma}(\mathbf{S})$ stems from Welch [15]. On the other extreme, identical rows give $\boldsymbol{\alpha}(\mathbf{S}) = 1$ and if in addition all the entries of $\mathbf{S}$ are equal then also $\boldsymbol{\beta}(\mathbf{S}) = \boldsymbol{\gamma}(\mathbf{S}) = 1$. Clearly, identical sign patterns are not adequate for the MWC since the channels produce the same measurements. Intuitively, the mixing patterns $p_i(t)$ should differ from each other as much as possible in order to provide additional information on the signal. For these reasons, the best quality is attained when $\boldsymbol{\alpha}(\mathbf{S}), \boldsymbol{\beta}(\mathbf{S}), \boldsymbol{\gamma}(\mathbf{S})$ are low. Notice again the analogous requirement for low coherence in standard CS.

The following theorem is our main contribution. It relates the quantities $\boldsymbol{\alpha}(\mathbf{S}), \boldsymbol{\beta}(\mathbf{S}), \boldsymbol{\gamma}(\mathbf{S})$ to the probability of satisfying the ExRIP.

*Theorem 2:* Let $\mathbf{\Phi} = \mathbf{SF}/\sqrt{mM}$ be the scaled sensing matrix of the MWC. If the nonzeros are drawn from a symmetric distribution, then $\mathbf{\Phi}$ has the ExRIP with probability at least

$$p = 1 - \frac{(1 - C_K)\rho_M (1 + \boldsymbol{\alpha}(S) - 2\boldsymbol{\beta}(S))}{\delta_K^2} \quad (14)$$
$$- \frac{(B_K - C_K)\rho_M (\boldsymbol{\gamma}(S) - \boldsymbol{\beta}(S)) + C_K M \boldsymbol{\beta}(S) - 1}{\delta_K^2}$$

where $C_K = \mathbb{E}\{\sum_{i=1}^{K} |u_i|^4/\|\mathbf{u}\|^4\}$, $B_K = \mathbb{E}\{|\sum_{i=1}^{K} u_i^2|^2/\|\mathbf{u}\|^4\}$ are distribution-dependent constants, $u_i$ are the $K$ random variables representing the nonzeros of $\mathbf{u}$, and $\rho_M = M/(M-1)$.

The symmetric distribution in the theorem refers to symmetry of the probability density function around the origin. For example, Gaussian, uniform and Bernoulli distributions satisfy this condition. Note that $B_K = 1$ whenever the nonzeros are real-valued. It can also be verified that $C_K = 3K/(2K + K^2)$ when $u_i$ are standard normal variables. Explicit formulas for other distributions exist, but we find it more convenient to sample the distribution and approximate $B_K, C_K$ by averaging. We note that the bound (14) may be negative when the ExRIP is not satisfied.

Theorem 2 is proved by calculating the second and the forth moments of $Z = \|\mathbf{\Phi u}\|/\|\mathbf{u}\|$, a somewhat tedious procedure. Point out that the $\sqrt{mM}$ scaling in Theorem 2 stems from the default scaling $1 \pm \delta_K$ in (5). We refer to [14] for the complete proof.

To realize the advantage of Theorem 2, we now evaluate the ExRIP for several choices of sign patterns. Not surprisingly, popular binary sequences that are used in communication channels appear adequate due to their low correlation-energy.

TABLE II: ExRIP guarantees for different sign patterns

| Family | Dimensions | | | Quality ×100 | | | ExRIP prob. $p$ | |
|---|---|---|---|---|---|---|---|---|
| | $m$ | $M$ | $2K$ | $\alpha(\mathbf{S})$ | $\beta(\mathbf{S})$ | $\gamma(\mathbf{S})$ | Normal | Uniform |
| Maximal | 80 | 511 | 24 | 1.438 | 0.196 | 0.408 | 0.932 | 0.931 |
| Gold | 80 | 511 | 24 | 1.255 | 0.198 | 0.199 | 0.939 | 0.939 |
| Hadamard | 80 | 512 | 24 | 1.250 | 1.094 | 1.238 | 0.000 | 0.000 |
| Random1 | 80 | 511 | 24 | 1.439 | 0.198 | 0.202 | 0.927 | 0.927 |
| Kasami | 16 | 255 | 12 | 6.667 | 0.392 | 0.294 | 0.689 | 0.675 |
| Random2 | 40 | 195 | 24 | 3.025 | 0.526 | 0.537 | 0.856 | 0.858 |

*B. Choosing the Sign Patterns*

The literature describes several families of binary sequences with proven correlation guarantees. The Maximal, Gold [4], Kasami [16], and Hadamard codes are families of sequences with different bounds on the self and the mutual correlations. The first three families require $M = 2^n - 1$ with different limitations on the possible order $n$ and on the number of sequences within the code. The $M$-square Hadamard matrix provides up to $M = 2^n$ sequences.

In Table II we calculate the probability $p$ of satisfying the ExRIP for these codes using (14). Refer to [14] for the exact details of constructing the sequences. As before, $\delta_{2K} = \sqrt{2} - 1$ is used. Empirically, we noticed that random instances of $\mathbf{S}$ perform well as demonstrated in the table. This is reasoned by the cumulative nature of the measures $\alpha(\mathbf{S}), \beta(\mathbf{S}), \gamma(\mathbf{S})$, which forgive local peaks and averages the total power of the correlation functions. Combining the fact that random instances are not limited to pre-specified lengths together with the ability to compute $\alpha(\mathbf{S}), \beta(\mathbf{S}), \gamma(\mathbf{S})$ efficiently allows for a wise selection of the patterns. The probabilities in Table II are computed for complex variables $u_i$, for which both the real and the imaginary parts are either normal or uniform on $[-0.5, 0.5]$. Complex-valued MMVs result naturally in the MWC system [1].

Since Gold codes do not exist for $M = 255$ we used $M = 511$ for comparison. Maximal, Gold and even random sequences all seem adequate. In contrast, Hadamard sequences yield a poor probability estimate since their cross-correlation functions have many peaks, as implied by $\beta(\mathbf{S}), \gamma(\mathbf{S})$, which are one order of magnitude above the other codes. The small set of Kasami sequences, which offers only 16 patterns, is considered optimal in communication, since their auto- and cross-correlations achieve the Welch bound [15]. In our quality measures Kasami sequences achieve poor scores since the aggregated energy is high due to many peaks (at the Welch bound level). In contrast, maximal and Gold sequences have a few high local peaks but their average energy is low.

In the technical report [14], we show that (14) can be well approximated by

$$p \approx 1 - \frac{1}{m\delta_K^2}, \quad (15)$$

for the Maximal and Gold codes, and for complex-valued signals. Fig. 2 plots the probability of satisfying ExRIP, as predicted by Theorem 2, for increasing number of sampling channels and (complex-valued) normally distributed nonzeros. We added the approximated bound (15) for comparison. Evidently, the conditioning of the MWC has two dominant factors: the number of sampling channels $m$ and the required conditioning level $\delta_K$.

To conclude we comment that Theorem 2 proves that the MWC has the ExRIP. Relating the conditioning probability to success recovery by basis pursuit requires additional steps as carried out in [13] or as proposed in [3].

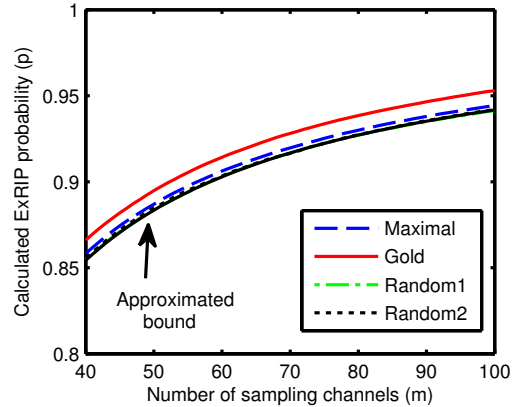

**Fig. 2:** The probability of satisfying the ExRIP depends mainly on the number of sampling channels. The black solid curve shows that the approximated bound (15) coincides with the theoretical curve of the *Random2* sequence family, whose dimensions match the MWC setting of [1].